\documentstyle[12pt]{article}
\begin{document}
\setlength{\baselineskip}{12pt}

\vspace*{36pt}
\noindent
{\bf WHAT IS ACHIEVED BY DECOHERENCE?}
\renewcommand{\thefootnote}{\fnsymbol{footnote}}
\footnote{ To appear in ``New Developments on Fundamental Problems in Quantum
Physics" (Oviedo II), M. Ferrero and A. van der Merwe, edts., (Kluwer Academic,
1997).}

\vspace{48pt}

\noindent \hspace*{1.7cm}{\bf H. D. Zeh}

\noindent \hspace*{1.7cm} {\it Institut f\"ur Theoretische Physik,
Universit\"at Heidelberg}

\noindent \hspace*{1.7cm}{\it Philosophenweg 19, D-69120 Heidelberg,
Germany}

\vspace{24pt}
\noindent A short critical review of the concept of decoherence, its
consequences, and its possible implications for the interpretation of quantum
mechanics is given.

\vspace{12pt}

\vspace{24pt}
\noindent {\bf 1. WHAT IS {\it MEANT} BY DECOHERENCE?}

\vspace{12pt}
\noindent There seems to be some confusion in the literature not only on what
may actually be achieved by decoherence, but also on how this concept has to be
defined. I will here ``consistently" use it in terms of wave functions
(not ``histories"), since state vectors represent the
established kinematical concept of quantum theory, while events (of which
histories are often assembled) will turn out to be derivable in a certain
sense in terms of wave functions just by means of decoherence.

Therefore, by decoherence I mean the practically irreversible and practically
unavoidable (in general approximate) disappearance of certain phase relations
from the states of local systems by interaction with their environment
according to the Schr\"odinger equation. Since phase relations cannot
absolutely disappear in a unitary evolution, this disappearance can only
represent a {\it delocalization}, which means that the phases ``are not there"
any more, neither in the system nor in the environment, although they still
exist in the total state that describes both of them in accordance with quantum
nonlocality.

This can also be described as the dynamical disappearance of
nondiagonal elements from the subsystem density matrices in a certain basis.
However, the concept of a density matrix would then first have to be defined
(or justified) in terms of the conceptually presumed wave functions. This does
not present a problem ``for all practical purposes" (FAPP), that is, when using
the ``rules" of quantum theory for calculating probabilities for outcomes of
measurements, for example in the form of expectation values
\begin{equation}
\langle A_{local} \rangle = Trace[A_{local}\rho_{total}] =
Trace[A_{local}\rho_{local}]
\label{prob}
\end{equation}
in a Hilbert space ${\cal H}_{total} = {\cal H}_{local}\bigotimes {\cal
H}_{env}$, where
$\rho_{local}=Trace_{env}[\rho_{total}]$, while
$\rho_{total}=
\vert
\psi_{total} \rangle \langle \psi_{total} \vert$ may represent a wave function.
Obviously, this procedure FAPP makes essential use of the probability
interpretation (albeit without using any assumptions about the
conceptual nature of the elements of the corresponding statistical ensembles,
or on how they may possibly arise dynamically). Conclusions derived in terms
of the density matrix should therefore not be used to refer to the probability
interpretation itself in order to avoid circular arguments.

Nonlocal phase relations are thus equivalent to entangled states of spatially
separated systems (quantum nonlocality). They can be conveniently described
by their Schmidt canonical form
\begin{equation}
\psi_{total}= \sum_{n}{\sqrt{p_n} \phi_n^{local}\Phi_n^{env} } \quad ,
\label{Schmidt}
\end{equation}
which is generically unique, and directly corresponds to the diagonal
representation of the density matrix for both subsystems, e.g.
\begin{equation}
\rho_{local}= \sum_n {\vert \phi_n^{local}\rangle p_n \langle
\phi_n^{local} \vert } \quad .
\end{equation}
This density matrix has a form {\it as though} it
represented an ensemble of wave functions $\phi_n^{local}$ with
respective probabilities
$p_n$. Equation (\ref {Schmidt}) demonstrates, however, that it does not. This
confusion between proper and ``improper" mixtures has given rise to the most
frequent misinterpretation of decoherence as leading to ensembles, and thus as
deriving the collapse of the wave function as a
stochastic process. Such a replacement of improper by proper
mixtures is clearly part of the {\it quantum state diffusion model} (insofar as
it goes beyond decoherence), and it is similarly used when ``selecting" {\it
consistent histories} from their superpositions (thus not from ensembles). Any
subsequent restriction ``FAPP only" will here solely refer to this ``as though"
--- not to the arising entanglement with the environment or its dynamical
consequences for the corresponding formal density matrix. Therefore, it does
neither mean ``wrong" nor ``useless".

For time-dependent states $\psi(t)$, their entanglement (\ref{Schmidt}) must in
general also depend on time. It may thus be entirely {\it caused} by
interaction, for example according to von Neumann's (unitary part of) ideal
measurements
\begin{equation}
( \sum_n {c_n \phi_n^{local} }) \Phi_0^{env}  \to \sum_n { c_n
\phi_n^{local} \Phi_n^{env} } \quad . \label{Neumann}
\end{equation}
This measurement-like process describes {\it pure decoherence} of the
system
in neglecting any re-action (or ``perturbation" by the environment).
Decoherence has therefore also been called ``continuous measurement"
(continuous in the sense of a permanent or unavoidable coupling).

The dynamical evolution described by Eq.\thinspace(\ref{Neumann}) contains an
obvious {\it arrow of time}, based on the belief that entanglement must be {\it
retarded}, that is, have causes in the past (``quantum causality"). Although
entanglement is not a {\it statistical} correlation (not based on
incomplete information), Eq.\thinspace(\ref{Neumann}) is analogous to
Boltzmann's assumption of molecular chaos, since it neglects any {\it
re}coherence (any local effects of the arising quantum correlations in the
future). In both cases, the existence of fluctuations demonstrates that this
time arrow is not a law, but a fact in accordance with time-symmetric laws.

Genuine measurements (for example, of the passage of a particle through a slit
of an interference device) are known to destroy interference for trivial
reasons: the events following the observed passages may be counted separately,
and may then simply be added. However, this is not required for
decoherence, since (a) there need not be definite passages, and (b) the state
of the environment need not correspond to a controllable ``pointer state", from
which information about the passage could be retrieved. For this reason one
should rather speak of a measurement-{\it like} interaction with the
environment.

Decoherence is thus a normal consequence of interacting quantum
mechanical systems. It can hardly be denied to occur --- but it cannot explain
anything that could not have been explained before. Remarkable is only its
quantitative (realistic) aspect that seems to have been overlooked for long.
Entanglement is the norm --- not the exception ---, and it must have important
consequences. In particular, all macroscopic systems are never approximately
isolated, and must thus not be described by wave packets obeying a
Schr\"odinger equation (not even when a WKB approximation applies).
Furthermore, since a macroscopic system together with its environment is even
more macroscopic, quantum theory can provide a consistent description only when
applied to the universe as a whole. That means, one has to assume
$\psi_{total} =
\psi_{universe}$ in order to get unambigious results.

\vspace{24pt}
\noindent {\bf 2. ELEMENTARY SYSTEMS UNDER DECOHERENCE}
\vspace{12pt}

\noindent In order to illustrate the consequences of realistically taking into
account the environment of quantum systems FAPP, this section lists some
important examples of decoherence.
Because of the rich literature on the subject I will here refrain from
citing individual papers, and instead refer only to our recent review
\cite{giu}, which has just appeared, and which contains a bibliography
considerably exceeding the length of this article. Further contributions --- in
particular some impressive work by the Madrid group(s) --- can be found in
these proceedings.

\vspace{15pt}
\noindent{\bf 2.1 Trajectories of Macroscopic Variables}

\vspace{12pt}
\noindent Interfering paths of
mass points are equivalent to spatial waves. In a two-slit experiment with
``bullets" (dust particles or even large mole\-cules), not only their passage
through the slits, but the whole path would unavoidably be measured under {\it
all} realistic circumstances. No superposition of different positions
of such objects can ``be there" and lead to interference in
the probability distributions for subsequent local measurements. In this
respect, classical objects resemble alpha-particles in a cloud chamber. The
resulting entanglement leads to a density matrix for their center of mass
motion {\it as though} it represented an ever-increasing ensemble of narrow
wave packets following slightly stochastic trajectories. This is dynamically
described by a master equation
\begin{equation}
i\frac{\partial \rho(x,x',t)}{\partial t} = \frac{1}{2m} \left(
\frac{\partial^2}{\partial {x'}^2}  - \frac{\partial^2}{\partial x^2} \right)
\rho - i\Lambda (x-x')^2 \rho \quad , \label{master}
\end{equation}
which (including its coefficient $\Lambda$) can thus be derived from a
universal Schr\"odinger equation. One does
not have to {\it postulate} a phenomenological semigroup to characterize open
systems.

Since this situation has been sufficiently discussed in the literature, let me
here only emphasize that the required ``measurements" must indeed be
practically irreversible in order to avoid the possibility of ``quantum
erasure" (restoration of interference). Such a restoration would require,
however, that every single scattered particle were completely recovered.
(Experiments demonstrating quantum erasure in certain microscopic systems do
not represent restoration of the {\it whole} initial superposition.) Every
classical phenomenon, even in ``reversible" classical mechanics, is based on
such irreversible processes, with a permanent production of entropy that is
macroscopically negligible but large in terms of bits.

\vspace{15pt}
\noindent {\bf 2.2 Molecular Configurations and Robust States}

\vspace{12pt}
\noindent Chiral molecules, such as sugar,
represent another simple example of systems under decoherence. They are
described by wave functions, but in contrast to the analogous spin-3
state of the ammonia molecule, for example, not by energy eigenstates. The
reason is that it is their chirality (not their parity) which is continuously
``measured" by the scattering of air molecules (for sugar under normal
conditions on a decoherence time scale of the order $10^{-9} sec$). The
measurement of parity in sugar, or the preparation of (very short-lived) sugar
molecules in their energy eigenstates, is therefore practically excluded, since
it would require an even stronger coupling to the measurement device. (The
molecule
$PH_3$, discussed at this conference by Gonzalo, forms an amazing
intermediate situation, with parity eigenstates expected to exist
under exceptional environmental conditions.)

A further dynamical consequence then holds that each individual molecule in a
bag of sugar must retain its chirality, while a parity state --- if it had come
into existence in a mysterious or expensive way --- would almost immediately
``collapse" into a local mixture of both chiralities with equal probabilities.
Parity is thus not conserved for sugar molecules.
The resulting mixture would also be diagonal in the parity
representation if the diagonal elements were {\it exactly} equal. However,
every {\it actually} resulting value of chirality would be permanent,
that is, it would always be confirmed when measured again.

This dynamical ``robustness" of the chiral {\it wave
function} seems to characterize what we call a ``real" (in the operational
sense) or ``classical" property --- just like the spot on a photographic plate,
or any other ``pointer state" of a measurement device. It also seems to be
essential for the physical realization of memory (such as in DNA, brains or
computers --- with the exception of quantum computers, which are
instead extremely vulnerable to decoherence, as was discussed here by Ekert).
Robustness is thereby compatible with a (regular) time dependence, as
exemplified by the wave packet describing the center of mass motion of a
bullet.

Chemists are used to
describing the motion of the nuclei in large mole\-cules classically (for
example by rigid configurations), while representing the electrons by wave
functions. In general, they attribute this asymmetry to a Born-Oppenheimer
approximation. This argument is definitely insufficient, since a
straight-forward application of this approximation to molecules would lead to
energy (and angular momentum) eigenstates, with vibrational and rotational
spectra as known for the hydrogen (or ammonia) molecule. The same
insufficient argument is now found in {\it quantum gravity}, where it is used
to justify classical spacetime by employing a BO approximation with respect to
the Planck mass (see Sect.~2.4).

Pseudo-classical behaviour can in both cases be explained by means of
decoherence again: the positions of nuclei are permanently measured by
scattering molecules or photons. But why only
the nuclei (or ions), and why not even they in very small molecules? The
answer is quantitative and based on a delicate balance between internal
dynamics and interaction with the environment, whereby the density of states
plays an essential role. (Much work on details remains to be done!) Depending
on the quantitative situation, one will either obtain an approximately unitary
evolution, a master equation (with time asymmetry arising from
quantum causality), or complete freezing of the motion (quantum Zeno effect).
The situation becomes simple again only for a free nucleus,
which is described by Eq.\thinspace(\ref{master}).

\vspace{15pt}
\noindent {\bf 2.3 Charge Superselection}

\vspace{12pt}
\noindent Gau\ss' law in the form $q =
\int {{\bf E} \cdot d{\bf S} }/4\pi
$ tells us that a local charge is correlated with its Coulomb
field at any distance. It is a matter of taste whether this correlation is
considered as kinematics, or as {\it caused} in the form of the retarded field
resulting from the charge in the past. Conceived quantum superpositions of
different charge values,
\begin{equation}
\sum_q{c_q \psi_q^{total} } = \sum_q {c_q \chi_q \Psi_q \lbrace field \rbrace}
=  \sum_q {c_q \chi_q \prod_r {\Psi_q \lbrace field(r) \rbrace} } \quad ,
\end{equation}
would therefore be nonlocal. Here, $\chi_q$ is the local charge state,
$\Psi_q\lbrace field \rbrace \allowbreak = \prod_r{\Psi_q \lbrace field(r)}
\rbrace$ a pseudo-classical field functional representing its Coulomb field,
symbolically written as a direct product of states on spheres with radius
$r$. The local charge system itself (possibly including its field within a
small sphere) would then be described by a density matrix of the form
\begin{equation}
\rho_{local} = \sum_q { \vert \chi_q \rangle \vert c_q\vert ^2 \langle \chi_q
\vert } \quad ,
\end{equation}
since the field states outside $r$ can be assumed to be mutually
orthogonal for different charges $q$. The charge is therefore decohered by its
own Coulomb field, and no superselection rule has to be {\it postulated}.

While this result is satisfactory for the theory, a more practical
question is at what distance, and on what time scale, two compensating charges
(or a charge in a superposition of different positions) are decohered (here
possibly in a reversible manner) by the resulting dipole field. Theoretical and
experimental contributions to an answer can be found in these proceedings (cf.
Sols or Hasselbach). Notice that decoherence will thus allow one to
distinguish between quantum field theory and action at a distance (where
decoherence would occur only after the field has reached and changed the state
of absorbing matter).

The gravitational field caused by any mass is analogous to the Coulomb field
of a charge. Superpositions of different masses should therefore be decohered
by the quantum state of spatial curvature. However, there is no elementary
mass, and nobody has as yet reliably estimated at what mass difference
decoherence by the correlated quantum state of curvature becomes essential.
Quite obviously, some superposition of different mass-energy remains
dynamically relevant in the form of (observed) time-dependent quantum states of
local systems.

\vspace{15pt}
\noindent {\bf 2.4 Classical Fields and Gravity}

\vspace{12pt}
\noindent Not only the quantum states of charged
particles are decohered by their fields, the quantum
field states are in general also decohered by source particles on which they
react. {\it Coherent states} (which represent classical fields) have in this
situation been shown to be robust in a similar way as chiral molecules or wave
packets for the positions of macroscopic objects. This is obviously the
reason why no superpositions of macroscopically different ``mean fields", or
different vacua (as they may arise through spontaneous symmetry breaking), have
ever been observed.

These arguments must as well apply to quantum gravity. One does not
have to know its precise form (for example, after its expected unification with
other forces) in order to conclude the existence of entangled superpositions of
matter and geometry (as far as this distinction remains valid). Therefore, the
classical appearance of spacetime geometry, with its lightcone
structure as presumed in conventional quantum field theory, is no reason not to
quantize gravity, since it should be explained by decoherence in the same way
as the classical appearance of a bullet or an electromagnetic field. The
resulting density matrix (functional) for the gravitational tensor field must
be expected to behave as though it represented a statistical mixture of
different classical curvature states (to which the observer is correlated ---
see Sect.~4). The beauty of Einstein's theory can hardly be ranked so much
higher than that of Maxwell's in order to justify its exemption from a well
established and general principle of physics (in particular, as this would lead
to incompatibilities with the uncertainty relations for matter --- known from
the Bohr-Einstein debate).

The
entropy characterizing a black hole (or the thermal field of an accelerating
Unruh detector) is thus due to a similar ``local perspective" of a strongly
entangled subsystem as the entropy produced according to the master equation
(\ref{master}) for a mass point. An event horizon
need not be different from any other pseudo-classical property, and even the
disappearance of information behind a classical horizon would not describe a
{\it real} collapse. For example, a succession of simultaneities
carrying global states which contain an Unruh detector along an external world
line may be chosen to remain all entirely outside the horizon (as
described by the Schwarzschild time coordinate).

\vspace{15pt}
\noindent {\bf 2.5 Quantum Jumps}

\vspace{12pt}
\noindent Exponential decay of excited states is usually regarded as
the standard manifestation of a fundamental quantum indeterminism. Stochastic
decay into a single channel would in fact lead to an exact exponentially
decreasing non-decay probability. However, the Schr\"odinger equation for a
mass point in a potential barrier, for example, leads to an {\it approximately}
exponential time-dependence {\it of the wave function}. It
represents the {\it superposition} (not an ensemble) of different decay times.
The corresponding interference is incompatible with exact exponential decay
even in free space, and it has been confirmed in reflecting cavities as
``coherent state vector revival".

On the other hand, if the decay fragments interact strongly
with the surrounding matter, any interference between ``decayed" and ``not yet
decayed" must practically irreversibly vanish on a very short decoherence time
scale at every moment of time. The corresponding apparent quantum jumps at
almost discrete times are observed as ``clicks" of a Geiger counter (in
analogy to pseudo-classical local ``spots" on a photographic plate). They have
also been seen with individual atoms coupled to laser fields as a sudden
appearance and disappearance of ``dark periods". For electrons tunneling from
the tip of a metal needle, decoherence between different emission times may
explain the short longitudinal coherence lengths reported at this conference by
Hasselbach (while the dipole moments discussed in Sect.~2.3 are in these
experiments represented by the {\it transversal} separation of the interfering
paths). A decaying system strongly interacting with its environment is thus
effectively described by a stochastic process with discrete decay times, and
the resulting time dependence will be almost exactly exponential (as long as
quantum causality governs decoherence).

It seems that this situation of continuously monitored decay has led
to the myth of quantum theory as a stochastic theory that
describes fundamental discontinuous quantum {\it events}. For
example, Bohr formulated in 1928 that ``the essence (of quantum theory) may be
expressed in the so-called quantum postulate, which attributes {\it to any
atomic process} an essential discontinuity, or rather individuality \dots" (my
italics). If this were correct, we would not even be able to describe lasers.
Heisenberg and Pauli similarly emphasized that their preference for matrix
mechanics originated in its (as it now seems
misleading) superiority in describing discontinuities. According to the
Schr\"odinger equation, the underlying entanglement processes occur smoothly
--- but recall now that all conclusions regarding ensembles are still based
FAPP on the yet unspecified probability interpretation (\ref{prob})!

\vspace{15pt}
\noindent {\bf 3. CONSEQUENCES FOR THE {\it MOTIVATION} OF VARIOUS
INTERPRETATIONS OF QUANTUM THEORY}
\vspace{12pt}

\noindent How has the {\it fundamental} probability interpretation, used in
Eq.\thinspace(\ref{prob}), then to be understood? And where precisely does it
have to be applied in view of the problem of where to position the
Heisenberg cut? The greatly differing opinions of physicists on this point are
surprising, and proof that the answer does not matter FAPP. Obviously,
decoherence cannot give an answer either, although it may help to recognize
some interpretations as being essentially motivated by prejudice or tradition.

Many seem to believe, in the tradition of Heisenberg, that Eq.\thinspace
(\ref{prob}) describes probabilities for {\it classical} quantities to ``enter
reality". However, we have seen that these classical quantities can be
described by robust wave packets (in a sense which comes close to
Schr\"odinger's original attempts, but overcomes the problem of their
dispersion by means of decoherence). It is then not at all surprising that
classical concepts, if nonetheless assumed to apply, have to be restricted
either in validity or in their observability in
order to avoid contradictions. There is a rich vocabulary
for de- or circumscribing the first possibility (uncertainty, complementarity,
new logic, negative probabilities, consistency etc.). Those of these
restrictions which are well defined can easily be {\it explained} in terms of
wave packets (for example, by means of the Fourier theorem). The second
possibility is represented by Bohm's ``surrealistic trajectories". As a
reflection on some contributions to this conference let me also emphasize that
it appears useless and even quite misleading to describe a certain subset of
experiments within classical concepts, while simply neglecting the crucial
rest.

A second class of interpretations suggests a description of reality in
terms of yet unknown (``hidden") variables. However, the evidence related to
Bell's inequalities clearly demonstrates (within limitations which must hold
for {\it all} empirical proof --- as known at least since David Hume, and as
some physicists now seem to be rediscovering in detail) that these new
variables would have to contain precisely the same nonlocality as
entangled wave functions on configuration space.

Then why reject the wave function itself as describing ``real" physical states?
I think that this natural position is now strongly supported by the
beautiful experiments with individual atoms in cavities (Schr\"odinger
cat states, quantum engineering, phase space tomography etc.), which were
discussed at this conference by several speakers. Thus, in a third class of
interpretations of the fundamental probabilities one considers jumps between
{\it wave functions} (usually described by a ``stochastic Schr\"odinger
equation"). However, why should the fundamental processes represented by (1)
then occur precisely where they {\it appear} to occur and are readily described
as decoherence?

\vspace{15pt}
\noindent {\bf 4. CONSEQUENCES OF DECOHERENCE FOR THE INTERPRETATION OF
QUANTUM MEASUREMENTS}
\vspace{12pt}

\noindent The probabilities described by Eq.\thinspace({\ref{prob}) are
usually understood as referring to {\it measurements}. In many situations we
may be satisfied just with entangled states, but individual measurements have
definite outcomes, while von Neumann's or any similar unitary interaction lead
to their superposition. This contrast defines what is usually called the {\it
measurement problem}. John Bell insisted that we should not speak of
measurements, but that would not resolve the dilemma. He obviously meant that
we should treat measurements (and the occurrence of events?) as a normal
physical process. That is precisely what many people are trying to do --- but
in terms of which kinematical concepts?

When describing macroscopic objects (such as measurement and registration
devices) FAPP by the apparent ensembles of wave functions resulting through
decoherence, we have in fact applied the probability interpretation
(\ref{prob}) to them. Completion of a measurement for this purpose thus
requires including the reading of the pointer state. The corresponding chain of
interactions would read
\begin{eqnarray}
 \left( \sum_n {c_n \phi_n }\right) \psi_0^{app} \chi_0^{env}
\Phi_0^{obs}
 & \buildrel \rm meas. \over \longrightarrow &
\left(\sum_n { c_n
\phi_n \psi_n^{app} } \right)  \chi_0^{env} \Phi_0^{obs} \nonumber\\
 \buildrel \rm decoh. \over \longrightarrow
\left( \sum_n { c_n \phi_n \psi_n^{app}   \chi_n^{env} }\right) \Phi_0^{obs}
& \buildrel \rm obs. \over \longrightarrow &
 \phi_{n_0} \psi_{n_0}^{app}   \chi_{n_0}^{env}  \Phi_{n_0}^{obs}
\end{eqnarray}
with probability $|c_{n_0}|^2$,
where the last step is itself a long and complex chain. Only when it has been
completed is the assumption of a collapse of the wave function without
prejudice empirically indicated. Probabilities can then only be understood as
representing frequencies
$f(n)$ in the results $n_{01},n_{02},\dots,n_{0N} $ of long series of $N$
equivalent measurements, described by  {\it one} final wave function
containing an observer state
$\Phi_{n_{01},n_{02},\dots,n_{0N}}^{obs}$ (cf. Mittelstaedt's contribution to
this volume).

If a real collapse occurred before the onset of decoherence, it would almost
certainly have directly observable consequences. If it occurred later in the
chain of interactions, decoherence could remain essential for the classical
appearance of the world. This situation would then lead to a ``partial Everett
interpretation", in which certain components of the wave function, although
they existed, would appear to be absent to an appropriate observer state. If
decoherence itself practically triggered a collapse (similar to the GRW model),
it would again be dangerously close to observation, since the environmental
situation may be altered to some extent, while fundamental dynamical terms had
to be fixed.

Thus recall von Neumann's motivation for the collapse: his aim was to
re-establish a psycho-physical parallelism based on definite states
$\Phi_{n_0}^{obs}$ of the observer (by whatever system an observer may be
physically represented). However, there {\it is} a definite state
$\Phi_n^{obs}$ in {\it each} robust component of a global superposition $
\sum{c_n \phi_{n}
\psi_{n}^{app}
\chi_{n}^{env}  \Phi_{n}^{obs} }$. Each component behaves then dynamically
{\it as though} it represented the complete and only world (even though this
behaviour is here derived from the assumption that it does {\it not}).

These
branching robust components describe microscopic histories which do not {\it
individually} determine their past. The essential reason for the dynamical
independence (robustness) of the arising branches in their future (the absence
of recoherence) is not simply the linearity of the Schr\"odinger equation, but
paradoxically the very same retarded nature of quantum entanglement (quantum
causality --- Eq.\thinspace(\ref{Neumann})) which seems to be responsible for
the consistency (hence existence) of documents about {\it macroscopic} history
(the ``fixed past").

According
to this picture it is the apparently observed quasi-classical world that exists
only FAPP ({\it viz.}, with respect to each
$\Phi_{n}^{obs}$ and his or her ``friends" $\Phi_{n}^{obs'}$).
The choice between the Everett interpretation and a vaguely located collapse
thus remains presently a matter of taste.

\vspace{12pt}
\noindent I wish to thank Erich Joos and Claus Kiefer for their comments.

\vspace{15pt}
\noindent {\bf REFERENCES}

\def\refname{}

\end{document}